\documentclass{jais}

\journal{Preprint}
\vol{October 2026}

\usepackage{amsmath}
\usepackage{amssymb}
\usepackage[hidelinks]{hyperref}

\renewcommand{\thefootnote}{\fnsymbol{footnote}}

\begin{document}

\title{Muography Simulations for Carbon Capture and Storage Monitoring}

\author{
Hamid Basiri,\auno{1}\footnotemark\
Andrea Giammanco,\auno{1}
Jun Matsushima,\auno{2}
Mohammed Y.~Ali,\auno{3}
Eduardo Cortina Gil,\auno{1}
Ahmed Eleslambouly,\auno{3}
and Hiroyuki K.M.~Tanaka\auno{4}
}

\address{
\begin{flushleft}
$^1$Centre for Cosmology, Particle Physics and Phenomenology (CP3), UCLouvain, Chemin du Cyclotron 2, Box L7.01.05, Louvain-la-Neuve, B-1348, Belgium\\
$^2$Department of Environment Systems, Graduate School of Frontier Sciences, The University of Tokyo, Tokyo, Japan\\
$^3$Department of Earth Sciences, Khalifa University of Science and Technology, Abu Dhabi, United Arab Emirates\\
$^4$Earthquake Research Institute, The University of Tokyo, Tokyo, Japan
\end{flushleft}
}

\begin{abstract}

\begin{flushleft}
Muography can monitor CO$_2$ storage through reservoir density changes, but site-scale simulations are costly: reservoirs can be deep and borehole detectors are small. We address both with two approaches. First, a two-stage workflow splits the kilometer-scale simulation: a fast code transports muons through the overburden, and a general-purpose Monte Carlo code through the reservoir and detectors. Second, a geometric filter at the source keeps only muons whose direction and position let them reach the detectors, within a margin for multiple scattering, and a range cut removes those with too little energy. Both are available in the UCMuon simulation suite for muon generation, filtering and transport. At a deep (800--1000~m) and a shallow (65~m) CO$_2$ storage site, each approach reduces the computing time about 700-fold, and the filtered simulation gives the same hit rate as a full one, as long as the margin exceeds the muons' sideways scattering drift.

\par\vspace{8pt}


\end{flushleft}
\end{abstract}

\maketitle

\renewcommand{\thefootnote}{\fnsymbol{footnote}}
\footnotetext{Corresponding author: hamid.basiri@uclouvain.be}

\section{Introduction}

Carbon Capture and Storage (CCS) plays an important role in reducing emissions from heavy industry and power generation~\cite{bui2018ccs}. In this technique, CO$_2$ is captured, compressed, and injected into deep saline aquifers or depleted hydrocarbon fields. Storage is only acceptable if operators can demonstrate that the CO$_2$ remains securely emplaced.

Time-lapse (4D) seismic surveying is the reference monitoring method, used alongside other techniques. While seismic surveys image the plume geometry well, they are periodic and costly~\cite{furre201720}, and they can be environmentally disruptive~\cite{nowacek2015marine}. Furthermore, seismic velocity reacts strongly to the first few percent of gas and much less to further saturation, so saturation and total mass are not uniquely determined~\cite{landro2001discrimination}. Gravity monitoring is directly sensitive to mass changes, but it integrates over a broad volume and has limited spatial resolution. Density logs measure the bulk density accurately, but only near the well and only when it is logged. Continuous monitoring of density changes in the formation would complement these techniques.

Muography has the potential to fill this gap. Cosmic-ray muons are available everywhere on the Earth's surface, and high-energy muons penetrate hundreds of meters of rock. Their survival to a given depth depends on the density integrated along their path~\cite{bonechi2020atmospheric}. When CO$_2$ replaces denser brine, the opacity (the density integrated along the muon path) decreases and the detected muon rate rises. The concept was proposed for CO$_2$ storage in 2012~\cite{kudryavtsev2012monitoring}, and simulation studies have since assessed its sensitivity, mainly with simplified geometries and large idealized detection planes~\cite{matsushima2025contributions}. Laboratory experiments have shown that muon attenuation recovers the density and saturation of porous samples~\cite{eleslambouly2025laboratory}. Compact portable detectors~\cite{chaiwongkhot2022portable} and borehole-compatible detectors~\cite{bonneville2017novel,derrico2023borehole} have made underground measurements practical, and borehole muography is now used in the field, for example in mineral exploration down to about 1~km depth~\cite{ideon2025caber}. At the Svelvik CO$_2$ Field Lab, Geoptic deployed two strings of 13 detectors during a shallow injection in 2023 and observed an uncorrected rate increase of about 4.5\% in a detector at 38~m depth~\cite{steer2024spe,thompson2024get}.

Site-scale simulations are indispensable for both survey design and data interpretation, but full Monte Carlo modeling is computationally prohibitive for two reasons. The first is the \emph{depth problem}: muons must be transported through hundreds of meters to kilometers of overburden. The second is the \emph{acceptance problem}: a borehole detector is only about 6~cm wide, so almost every muon generated at the surface misses it. This contribution describes how we address both problems, with a deep example (Sleipner), where CO$_2$ is supercritical, and a shallow one (Svelvik), where it is a gas.

\begin{figure*}[!t]
\centering
\begin{minipage}[c]{0.686\textwidth}\centering
\includegraphics[width=\linewidth]{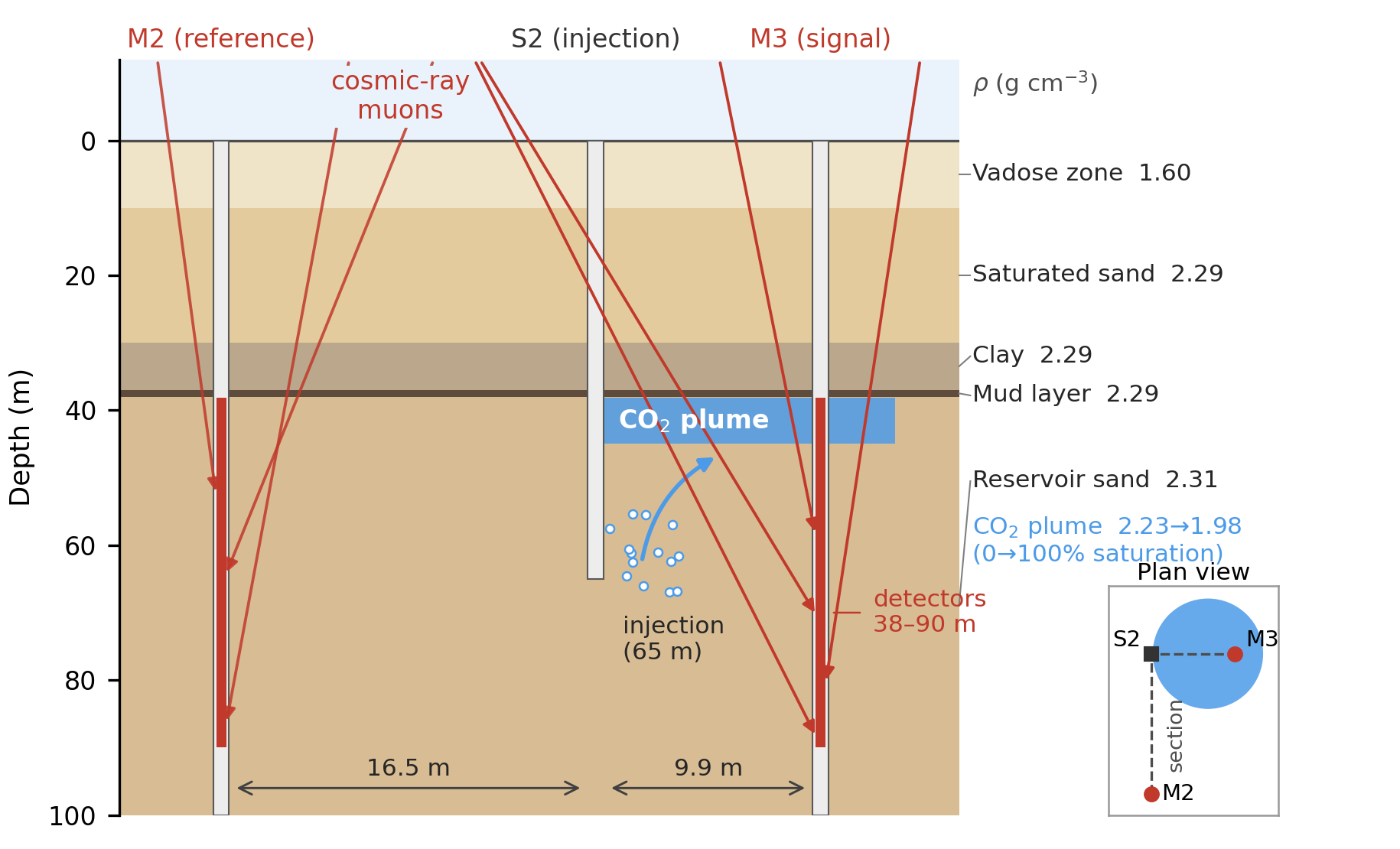}\\{\footnotesize (a)}
\end{minipage}\hspace{0.03\textwidth}
\begin{minipage}[c]{0.234\textwidth}\centering
\includegraphics[width=\linewidth]{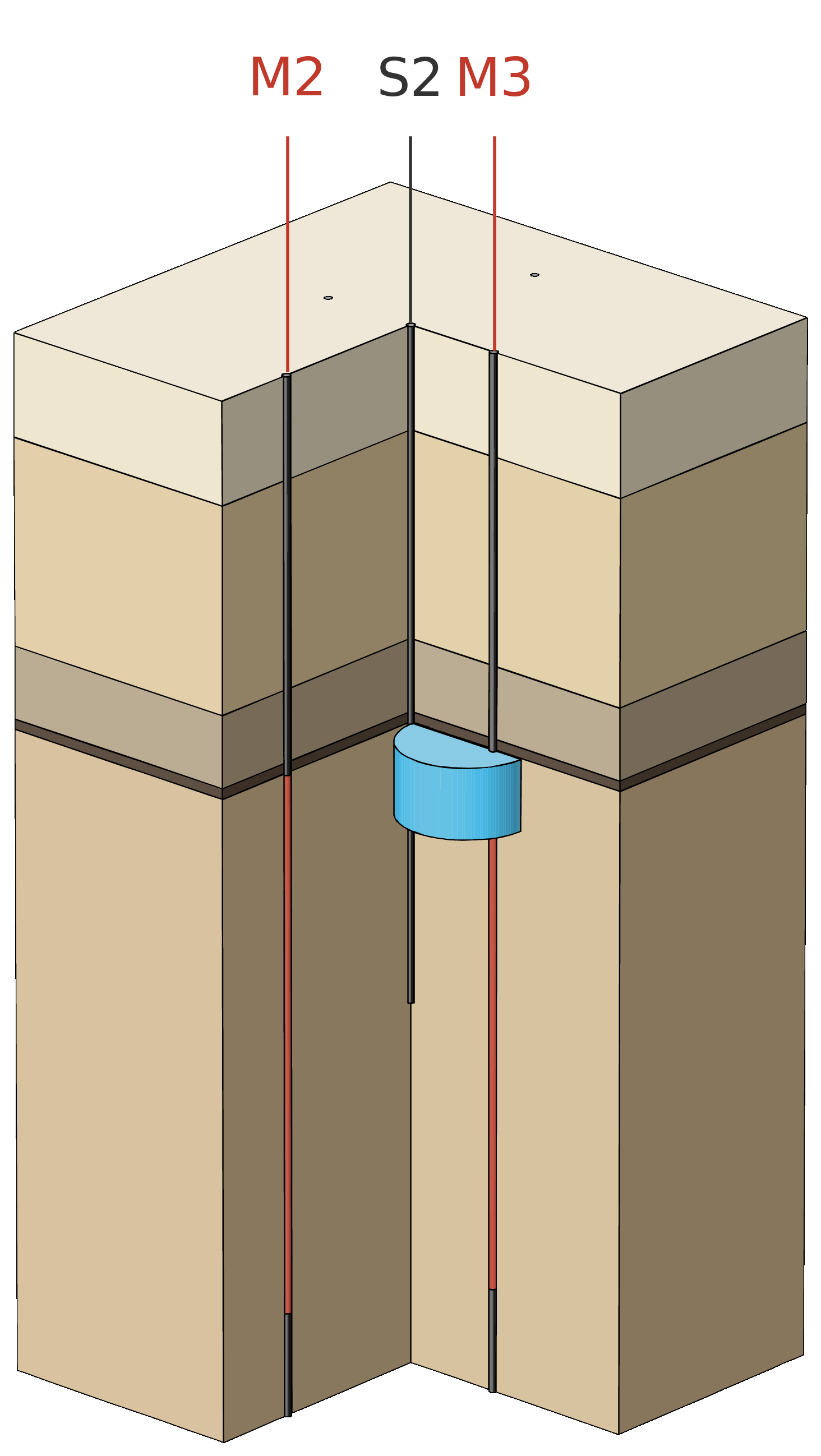}\\{\footnotesize (b)}
\end{minipage}
\caption{Simplified Svelvik model. (a) Measurement concept. CO$_2$ injected at 65~m in well S2 migrates upward and accumulates below the mud layer. Muons reaching the detector string in M3 (signal) cross the plume; borehole M2 serves as reference. The section runs from M2 to S2 and turns by 90$^\circ$ towards M3 (inset). Densities in g\,cm$^{-3}$; horizontal scale exaggerated $\approx$3.4$\times$, well diameters not to scale. (b) PHITS 3D view of the same model with one quadrant cut away, showing the wells, the detector strings (red) and the plume (blue); well diameters enlarged.}
\label{fig:concept}
\end{figure*}

\section{Methodology}

\subsection{Two-stage transport (depth problem)}

Fast muon transport codes such as MUSIC~\cite{kudryavtsev2009music} propagate muons efficiently through thick homogeneous media but cannot describe heterogeneous geometry. We therefore split the transport at a plane above the reservoir: the fast code carries the muons through the overburden, and their positions, directions and energies at the plane become the source of a general-purpose Monte Carlo code, here PHITS~\cite{sato2024phits}, which simulates the reservoir and the detectors. The overburden stage can be computed once and reused for every scenario. At the Sleipner site, where CO$_2$ is stored in a supercritical state (about 700~kg\,m$^{-3}$, compared with about 1020~kg\,m$^{-3}$ for the brine) in nine thin layers of the Utsira Sand at about 800--1000~m below sea level, $10^7$ vertical muons sampled from the CosmoALEPH sea-level spectrum~\cite{schmelling2013spectrum} were transported with MUSIC through 80~m of sea water and 700~m of rock, and PHITS propagated the 48.8\% that survived through the nine plume layers. This reduced the computational time by a factor of about 700 compared with transporting the full overburden and reservoir geometry in a single PHITS simulation~\cite{basiri2026investigation}.

\subsection{Geometric detector filter (acceptance problem)}\label{sec:filter}

At a shallow site the overburden is thin, and the cost is set by the detectors rather than by the depth: a borehole detector is only a few centimeters wide, so almost every muon generated over a source surface large enough to include inclined paths misses it. We therefore keep, before transport, only the muons that can reach a detector. The surface muons are generated with UCMuon, an open-source Fortran/Python simulation suite developed in our group~\cite{basiri2026ucmuon}. UCMuon draws the momentum, direction and position of each muon jointly from the surface spectrum and keeps it only if its straight-line track crosses a target volume around a detector. Because the generator counts every trial, the kept sample corresponds to an exact open-sky exposure: the number of trials divided by the muon rate through the source surface. A second, conservative cut removes kept muons whose energy is below 60\% of the continuous-slowing-down (CSDA) energy needed to reach the target along their track; such muons cannot reach the detectors. The remaining muons are written as a source for the transport code (PHITS in this work) and transported with unchanged physics. Because real muons scatter in the rock, the target must be larger than the detector by a margin that exceeds the expected lateral drift; otherwise muons that would scatter onto the detector are never generated.

\subsection{Svelvik site model}

The Svelvik CO$_2$ Field Lab is a SINTEF test site with an injection well reaching 65~m depth and four 100~m observation wells~\cite{jordan2022co2,jordan2024sparse}; the two used here, M3 and M2, lie 9.9~m and 16.5~m from the injection well S2, in perpendicular directions. Our simplified PHITS model (Fig.~\ref{fig:concept}) is a layered sediment column with bulk densities of 2.29~g\,cm$^{-3}$ from 10 to 38~m and 2.31~g\,cm$^{-3}$ below, assuming brine-saturated quartz sand (grain density 2.65~g\,cm$^{-3}$) with a porosity of about 0.2; the vadose zone above 10~m is set to 1.60~g\,cm$^{-3}$. The CO$_2$ plume is idealized as a cylinder of 6.5~m radius at 38--45~m depth, centered 3.2~m from borehole M3. With a porosity of 0.25, replacing all the brine by gaseous CO$_2$ (about 9~kg\,m$^{-3}$ at this depth) lowers the plume density from 2.23 to 1.98~g\,cm$^{-3}$; we compare this fully saturated case, an upper bound, with the same model without CO$_2$. Plastic-scintillator columns 3~cm in radius, spanning 38--90~m depth, sit in borehole M3, which passes through the plume, and in the reference borehole M2, which hardly sees it. To gain statistics, hits are scored on virtual cylinders of 0.5~m radius around the borehole axes, which have about 17 times the side area of the 3~cm columns; entries through their top face are not counted.

Muons were generated with the Guan spectrum~\cite{guan2015parametrization} between 15 and 1500~GeV, which covers the 15--100~GeV range of most muons reaching the Svelvik detectors. The lower energy limit lies below the $\approx$17~GeV (CSDA) needed to reach the top of the detectors. Zenith angles were limited to below $85^\circ$, and muons were started on a 500~m-radius surface disk, through which the Guan rate is $9.27\times10^6$~s$^{-1}$. PHITS transports the muons with ionization losses, energy straggling and multiple scattering; radiative processes and secondary particles are switched off, which changes the energy loss by about 10\% or less below 100~GeV and cancels to first order in relative signals.

\section{Results}

\subsection{Filter efficiency}\label{sec:filtereff}

\begin{figure}[!htb]
\centering
\includegraphics[width=\linewidth]{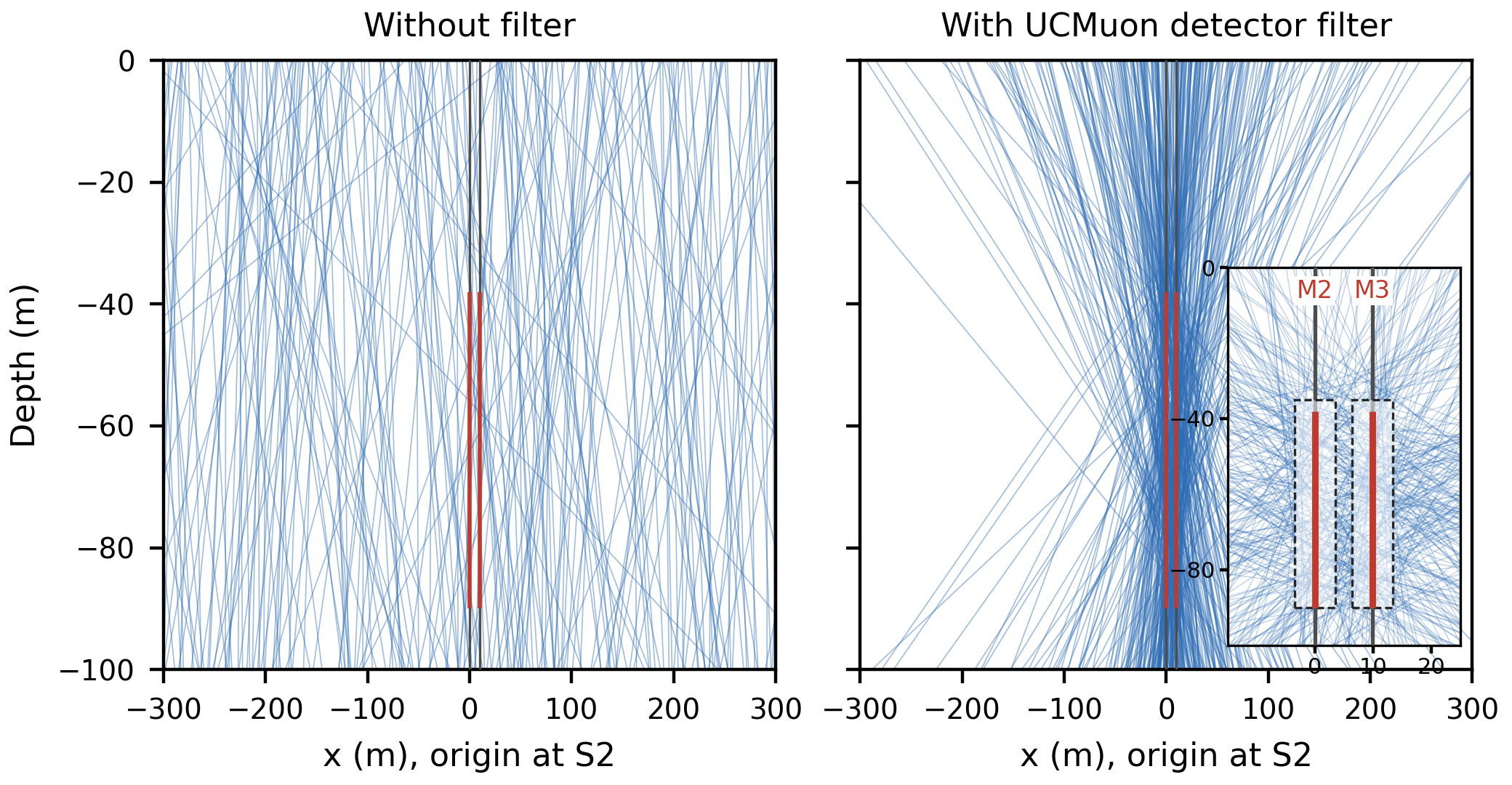}
\caption{Straight-line tracks of 300 surface muons in the Svelvik model, side view along $y$: $x=0$ at the injection well S2, M2 lies 16.5~m behind S2 and M3 at $x=9.9$~m. Left: unfiltered muons from the 500~m source disk. Right: muons kept by the UCMuon detector filter; the inset shows the two filter targets (dashed, 3.5~m radius, 35--90~m depth) around M2 and M3. Red: detectors (38--90~m).}
\label{fig:filtertracks}
\end{figure}

In the Svelvik model, the filter uses two targets, one around each instrumented borehole (M3 and M2): vertical cylinders of 3.5~m radius, the 0.5~m scoring cylinder plus a 3~m scattering margin (Sect.~\ref{sec:margin}), from 35 to 90~m depth. Of the muons generated on the 500~m-radius source disk, only 0.12\% have a straight-line track that crosses either target, so the filter removes more than 99.8\% of the histories the transport code would otherwise have to follow (Fig.~\ref{fig:filtertracks}). A sample of $10^8$ kept muons (before the range cut) corresponds to an open-sky exposure of 2.5~h. To measure the gain directly, we generated $1.98\times10^7$ surface muons, 2.1~s of open-sky exposure, and transported them through the Svelvik model both without the filter and after it (Table~\ref{tab:filter}). Per unit of exposure, the filter reduces the computing time by a factor of about 700, and about 1300 together with the range cut, which removes half of the kept muons. The hit rates in both boreholes agree within their statistical uncertainties of about 8\%.

\begin{table}[!htb]
\tbl{Filtered and unfiltered PHITS simulations of the same UCMuon surface muons in the Svelvik model (one CPU core, PHITS 3.36). Muons: number transported. CPU times and hit rates are per second of open-sky exposure; hits are entries into the 0.5~m virtual cylinders around M3 and M2.\label{tab:filter}}{%
\setlength{\tabcolsep}{4pt}
\begin{tabular}{lcccc}
\toprule
Source & Muons & CPU (s) & M3 hits & M2 hits \\
\colrule
Unfiltered               & $1.98\times10^7$ & 16\,800 & $82\pm6$ & $84\pm6$ \\
Detector filter          & 24\,000          & 23.4    & $73\pm6$ & $88\pm6$ \\
Filter + range cut       & 11\,758          & 12.7    & $80\pm6$ & $82\pm6$ \\
\botrule
\end{tabular}}
\end{table}

\subsection{Multiple scattering and the filter margin}\label{sec:margin}

\begin{figure}[!htb]
\centering
\includegraphics[width=0.75\linewidth]{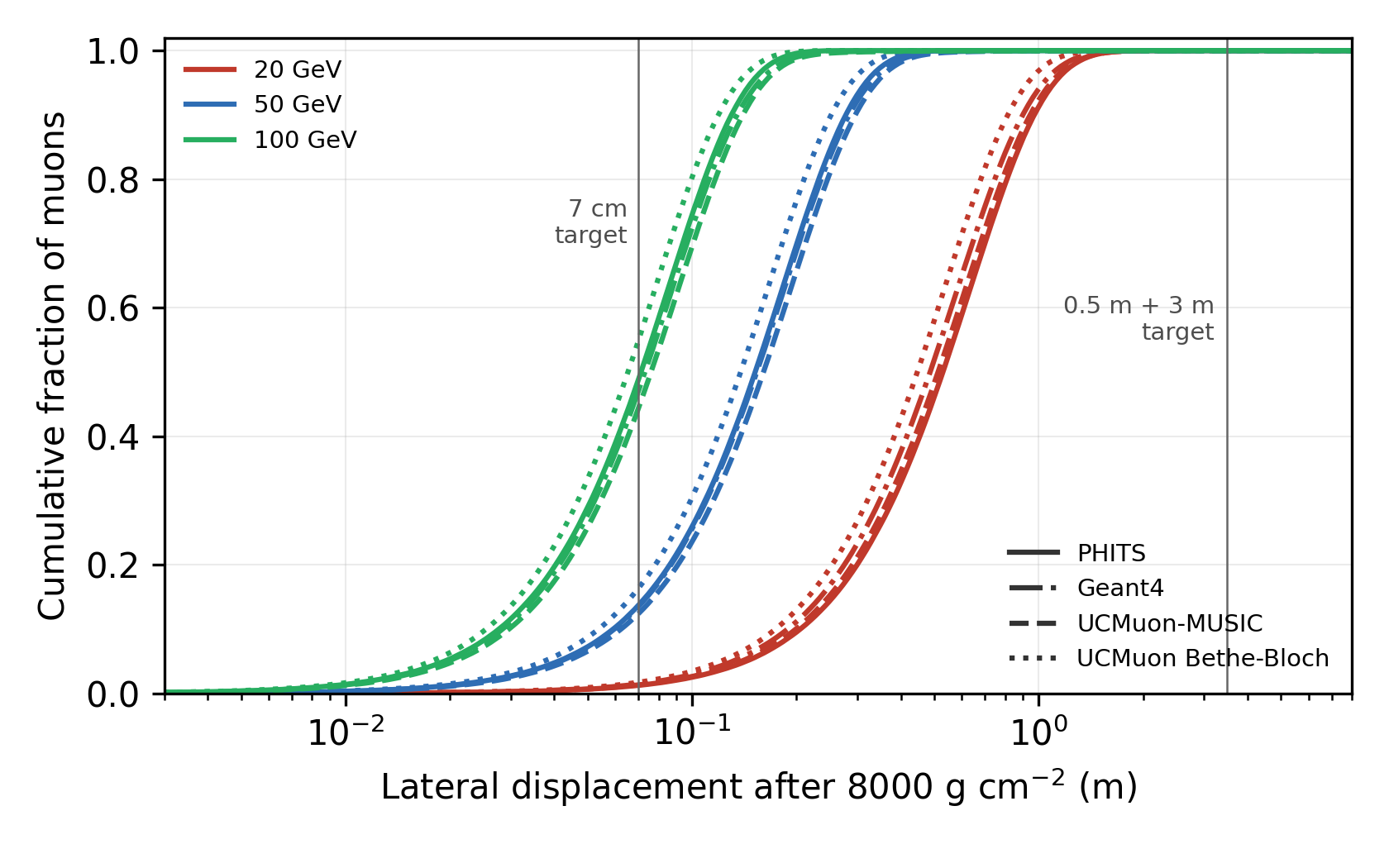}
\caption{Fraction of vertical 20, 50 and 100~GeV muons within a given lateral distance of their incident line after 8000~g\,cm$^{-2}$ of rock, which sets the filter margin; PHITS (solid), Geant4 (dash-dotted), UCMuon-MUSIC (dashed), UCMuon Bethe--Bloch (dotted), $10^6$ muons each. Left line: 7~cm target of the first runs; right line: 3.5~m production target (0.5~m scoring cylinder + 3~m margin). Deeper and inclined muons drift further (see text).}
\label{fig:drift}
\end{figure}

To quantify how far muons drift and to set the filter margin, we sent $10^6$ vertical muons at 20, 50 and 100~GeV through 8000~g\,cm$^{-2}$ of rock (40~m at 2.0~g\,cm$^{-3}$; SiO$_2$ in PHITS and Geant4, standard rock in UCMuon) with PHITS, with Geant4~\cite{agostinelli2003geant4,allison2016recent} using its FTFP\_BERT reference physics list, and with two UCMuon engines: MUSIC, and a Bethe--Bloch engine with Highland multiple scattering (Table~\ref{tab:drift}, Fig.~\ref{fig:drift}). PHITS, Geant4 and UCMuon-MUSIC agree within about 10\% in the rms lateral displacement, which is 0.59--0.64~m at 20~GeV, 0.18--0.20~m at 50~GeV and 0.09--0.10~m at 100~GeV. The Highland-based engine gives a 10--20\% narrower distribution. At 20~GeV, close to the end of the range (8000~g\,cm$^{-2}$ is the CSDA range of a 17~GeV muon), survival is 96\% in Geant4 and MUSIC but 86\% in PHITS, which with the settings used here stops muons about 14\% earlier than CSDA range tables~\cite{groom2001muon}; this lowers absolute PHITS rates but cancels to first order in relative signals.

\begin{table}[!htb]
\tbl{Lateral drift of $10^6$ vertical muons after 8000~g\,cm$^{-2}$ of rock: fraction that survives, rms lateral displacement, displacement not exceeded by 99.9\% of the muons, and fraction of surviving muons displaced by less than 7~cm and 3.5~m from the incident line. Geant4: FTFP\_BERT; UCMuon-BB: UCMuon Bethe--Bloch engine.\label{tab:drift}}{%
\setlength{\tabcolsep}{2.8pt}
\begin{tabular}{llccccc}
\toprule
$E$ & Code & Surv. & rms & $q_{99.9}$ & $<$7~cm & $<$3.5~m \\
(GeV) & & (frac.) & (m) & (m) & (frac.) & (frac.) \\
\colrule
20  & PHITS              & 0.861 & 0.64 & 1.72 & 0.012 & 1.000 \\
    & Geant4             & 0.961 & 0.59 & 1.67 & 0.014 & 1.000 \\
    & UCMuon-MUSIC       & 0.956 & 0.62 & 1.77 & 0.013 & 1.000 \\
    & UCMuon-BB         & 1.000 & 0.54 & 1.42 & 0.017 & 1.000 \\
50  & PHITS              & 1.000 & 0.18 & 0.49 & 0.137 & 1.000 \\
    & Geant4             & 0.999 & 0.19 & 0.63 & 0.135 & 1.000 \\
    & UCMuon-MUSIC       & 0.999 & 0.20 & 0.74 & 0.124 & 1.000 \\
    & UCMuon-BB         & 1.000 & 0.17 & 0.44 & 0.163 & 1.000 \\
100 & PHITS              & 1.000 & 0.09 & 0.23 & 0.488 & 1.000 \\
    & Geant4             & 0.999 & 0.09 & 0.35 & 0.471 & 1.000 \\
    & UCMuon-MUSIC       & 1.000 & 0.10 & 0.43 & 0.444 & 1.000 \\
    & UCMuon-BB         & 1.000 & 0.08 & 0.21 & 0.549 & 1.000 \\
\botrule
\end{tabular}}
\end{table}

\begin{figure*}[!htb]
\centering
\includegraphics[width=0.95\textwidth]{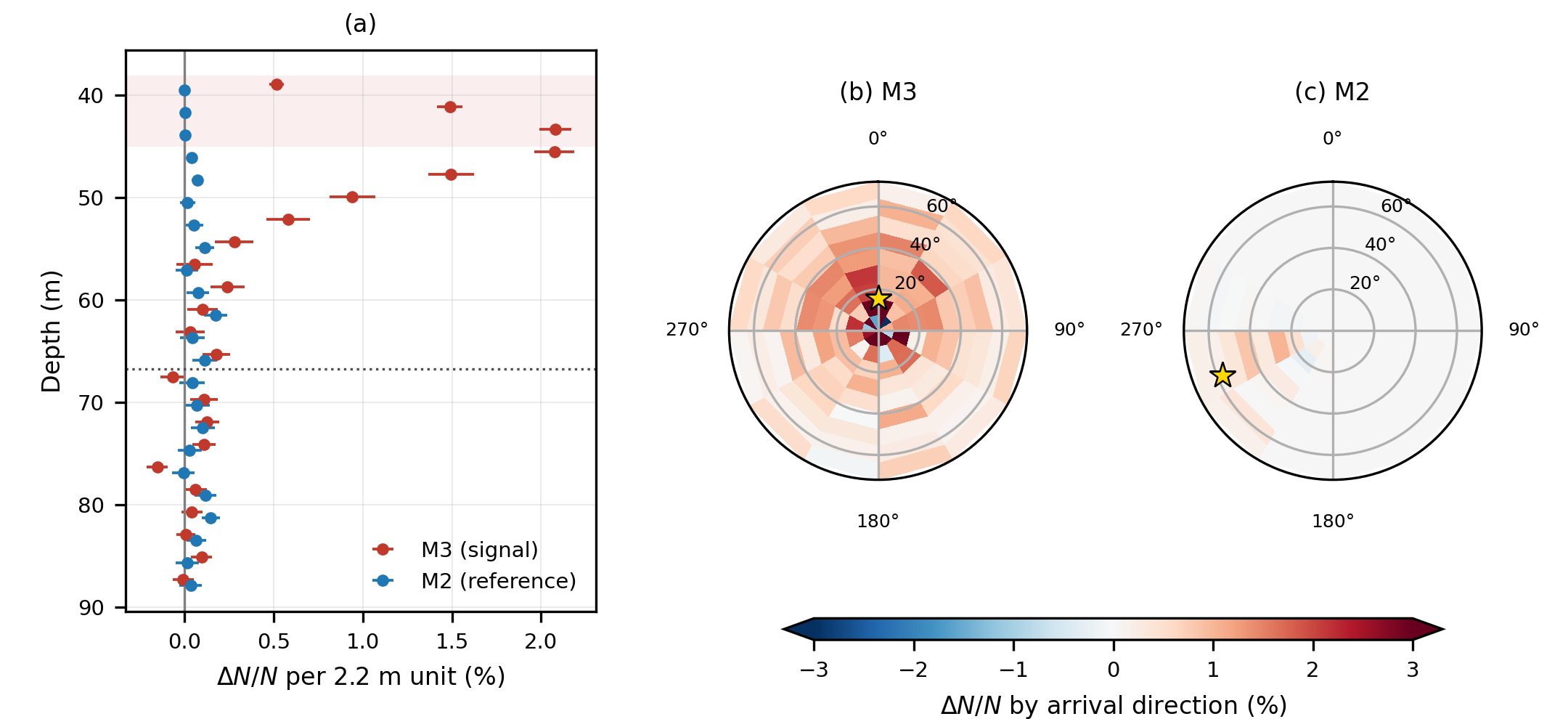}
\caption{Plume signal in the simplified Svelvik model: fully saturated plume versus no CO$_2$, $4.9\times10^7$ muons after the filter and range cut (2.5~h of open-sky exposure). (a) Relative change in the number of muons per 2.2~m unit in M3 (signal) and M2 (reference); error bars: Monte Carlo uncertainty of the paired runs; shaded band: plume depth; dotted line: end of a 13-unit string. (b, c) Relative change by arrival direction at M3 and M2 (zenith angle radially, azimuth in the model frame); the star marks the direction of the plume centre.}
\label{fig:signal}
\end{figure*}

The drift sets the filter margin. A target only a few centimeters larger than a 3~cm detector retains just 1--2\% of the 20~GeV muons, 12--16\% at 50~GeV and 44--55\% at 100~GeV, and the muons that would scatter onto the detector are never generated (Sect.~\ref{sec:filter}), so the hit rate is underestimated without warning. In our first Svelvik runs with a 7~cm target, PHITS recorded about five times fewer hits than straight-line tracing with the PHITS range predicted. With a 0.5~m target and a 3~m margin, all vertical test muons stay within the target in every code (99.9\% within 1.8~m). The margin is larger because the drift grows with depth, inclination and lower energy threshold: in the Svelvik production geometry, which includes all arrival directions, 99\% of the hits lie within about 3~m of their straight-line track, and the PHITS hit rate agrees with straight-line tracing to within 2\%, as expected when as many muons scatter onto the detector as off it. As the acceptance of a long vertical target grows linearly with its radius, the margin keeps about eight times more muons than the 0.5~m scoring cylinder alone, a cost already included in the gain of about 700 (Sect.~\ref{sec:filtereff}).

\subsection{Plume signal in the simplified model}

With this set-up, we transported $4.9\times10^7$ muons that passed the filter and the range cut, 2.5~h of open-sky exposure, through the model with and without CO$_2$. Both simulations use the same muon sample and random-number seeds, so their difference has a small Monte Carlo uncertainty: for the 13-unit string at M3 it is 0.02\%, nine times smaller than for independent runs, which would need about 80 times more muons for the same precision. Figure~\ref{fig:signal}(a) shows the change in the number of muons per 2.2~m unit, the length of the Geoptic detectors. At M3 it reaches 2.1\% in the units at the base of the plume and fades below 55~m: a vertical column is crossed mainly by inclined muons, and those reaching the deeper units pass beside the plume. Summed over a 13-unit string from 38 to 67~m, the count at M3 rises by $(0.94\pm0.02)\%$, while at M2 the change stays below 0.1\%. The arrival directions of the additional muons point to the plume from both boreholes (Fig.~\ref{fig:signal}(b, c)). A real 3~cm string in M3 records about 3.5 muons per second (surface flux without topography or buildings), so this change would reach $3\sigma$ after about 0.7 days of baseline and 0.7 days of monitoring, counting statistics only. Atmospheric variations of a few percent~\cite{steer2024spe} are much larger and must be removed, for example with the reference borehole. These numbers apply to an idealized, fully saturated plume and are not a prediction for the 2023 injection at Svelvik. Realistic plume models are left to a dedicated study.

\section{Conclusion}

Site-scale muography simulations for CCS face a depth problem and an acceptance problem. A two-stage MUSIC$\rightarrow$PHITS transport handles thick homogeneous overburdens; at Sleipner it gave a 700-fold speed-up. A geometric detector filter in UCMuon handles small borehole detectors; at Svelvik it reduces the computing time per unit of exposure by about 700 times, and 1300 times with a range cut, with the same hit rate within statistics. The filter's margin must exceed the multiple-scattering drift of the muons, here 0.2--0.6~m after 40~m of rock, otherwise the simulated hit rate is several times too low. In a simplified Svelvik model, a detector string that starts at the top of an idealized, fully saturated plume records about 1\% more muons, while the count in a reference borehole changes by less than 0.1\%. The two tools can be chained for deep sites monitored from boreholes, and both are available in UCMuon~\cite{basiri2026ucmuon}. Future work will apply the two-stage transport and the UCMuon filter, with a general-purpose code such as PHITS or Geant4, to realistic site-specific plume geometries and time-dependent CO$_2$ distributions.

\par\vspace{12pt}
\noindent{\bfseries\large Conflict of Interest\par}\nopagebreak
\vspace{4pt}\nopagebreak
\noindent
The authors declare that there is no conflict of interest regarding the publication of this paper.

\par\vspace{12pt}
\noindent{\bfseries\large Use of AI Tools\par}\nopagebreak
\vspace{4pt}\nopagebreak
\noindent
The authors used Claude (Anthropic), a generative AI tool, to assist with analysis and simulation scripts, consistency checks of the reported numbers against the simulation outputs, and language editing. The authors reviewed and verified all content and take full responsibility for the manuscript.

\par\vspace{12pt}
\noindent{\bfseries\large Acknowledgment\par}\nopagebreak
\vspace{4pt}\nopagebreak
\noindent
We thank Michael Jordan and Marcin Duda (SINTEF) for information on the Svelvik CO$_2$ Field Lab.
H.B. acknowledges support from a Fonds Sp\'ecial de Recherche (FSR) scholarship of UCLouvain.
A.G. is supported by the Fonds de la Recherche Scientifique -- FNRS under Grant No.~T.0129.25.
J.M. is supported by the Mohammed bin Salman Center for Future Science and Technology for Saudi Arabia--Japan Vision 2030 at the University of Tokyo (MbSC2030) and by the Earthquake Research Institute, the University of Tokyo, under the Joint Usage Research Program ERI JURP 2024-H-03.
Computational resources have been provided by the Consortium des \'Equipements de Calcul Intensif (C\'ECI), funded by the Fonds de la Recherche Scientifique de Belgique (F.R.S.-FNRS) under Grant No.~2.5020.11 and by the Walloon Region.

\bibliographystyle{JHEP}
\bibliography{ref}

\end{document}